\documentclass[prd,
twocolumn,
superscriptaddress,nofootinbib,%
tightenlines,showpacs]{revtex4}
\usepackage{mathrsfs}
\usepackage{latexsym,bm}
\usepackage{CJK}
\usepackage{graphicx}
\usepackage{indentfirst}
\usepackage{slashed}
\usepackage{amsmath}
\usepackage{amssymb}
\usepackage{color}
\usepackage{hyperref}
\usepackage{epsfig}
\hypersetup{CJKbookmarks=true}
\usepackage[titletoc]{appendix}
\usepackage{tabularx}
\usepackage{lipsum}

\newcommand{\be}{\begin{equation}} \newcommand{\ee}{\end{equation}}
\newcommand{\ba}{\begin{array}{c}} \newcommand{\ea}{\end{array}}
\newcommand{\bea}{\begin{eqnarray}} \newcommand{\eea}{\end{eqnarray}}

\newcommand{\MM}{\mathcal{M}}
\newcommand{\hco}{\hat{c}_1}
\newcommand{\pos}{{\sigma}_{17}\,}
\newcommand{\mos}{{\delta}_{17}\,}

\newcommand{\IBP}{\overline{\mathcal{I}}_{B\phi}}
\newcommand{\IB}{\overline{\mathcal{I}}_{B}}
\newcommand{\IP}{\overline{\mathcal{I}}_{\phi}}

\allowdisplaybreaks

\begin{document}
\title{\Large 
Masses and sigma terms of doubly charmed baryons up to $\mathcal{O}(p^4)$ \\in manifestly Lorentz-invariant baryon chiral perturbation theory}
\author{De-Liang~Yao}
\email{Deliang.Yao@ific.uv.es}
\affiliation{Instituto de F\'{\i}sica Corpuscular (centro mixto CSIC-UV), Institutos de Investigaci\'on de Paterna,
Apartado 22085, 46071, Valencia, Spain}

\begin{abstract}
 We calculate the masses and sigma terms of the doubly charmed baryons up to next-to-next-to-next-to-leading order (i.e., $\mathcal{O}(p^4)$) in a covariant baryon chiral perturbation theory  by using the extended-on-mass-shell renormalization scheme. Their expressions both in infinite and finite volumes are provided for chiral extrapolation in lattice QCD. As a first application, our chiral results of the masses are confronted with the existing lattice QCD data in the presence of finite volume corrections. Up to $\mathcal{O}(p^3)$ all relevant low energy constants can be well determined. As a consequence, we obtain the physical values for the masses of $\Xi_{cc}$ and $\Omega_{cc}$ baryons by extrapolating to the physical limit. Our determination of the $\Xi_{cc}$ mass is consistent with the recent experimental value by LHCb collaboration, however, larger than the one by SELEX collaboration. In addition, we predict the pion-baryon and strangeness-baryon sigma terms, as well as the mass splitting between the $\Xi_{cc}$ and $\Omega_{cc}$ states. Their quark mass dependences are also discussed. The numerical procedure can be applied to the chiral results of $\mathcal{O}(p^4)$ order, where more unknown constants are involved, when more data are available for unphysical pion masses.
\end{abstract}
\maketitle

\section{Introduction}
A doubly charmed baryon termed as $\Xi_{cc}^+$ was first reported by SELEX collaboration~\cite{Mattson:2002vu} and its mass was observed to be $3519\pm2$~MeV~\cite{Ocherashvili:2004hi}. Unfortunately, for very long time this state was not confirmed by any other experimental collaborations: 
FOCUS~\cite{Ratti:2003ez}, Babar~\cite{Aubert:2006qw}, Belle~\cite{Chistov:2006zj} or LHCb~\cite{Aaij:2013voa}. Very recently, renewed interest has been triggered in studying doubly charmed baryons due to the confirmation of the existence of the doubly charged state $\Xi_{cc}^{++}$ with a mass of $3621.4\pm0.78$~MeV by LHCb collaboration~\cite{Aaij:2017ueg}. Relevant theoretical efforts have been accumulated rapidly, for instance, in the investigations of their magnetic moments~\cite{Li:2017cfz}, weak decays~\cite{Wang:2017mqp,Wang:2017azm}, strong and radiative decays~\cite{Li:2017pxa,Xiao:2017udy},  interactions with light states~\cite{Guo:2017vcf}, {\it etc.} 

The masses of the doubly charmed baryons are basic quantities classifying the baryon spectrum.  Understanding the origin of the masses of ground-state baryons is one of the most important issues in hadron physics.  Especially for the $\Xi_{cc}$ baryons, the difference between the reported values of the masses by SELEX and LHCb collaborations are abnormally large, which is in conflict with the fact that the isospin breaking effect should be small as it is proportional to the mass difference of the $u$ and $d$ quarks. More specifically, the isospin splitting in $\Xi_{cc}$ baryons is estimated to be $m(\Xi_{cc}^{++})-m(\Xi_{cc}^{+})=1.41\pm0.12^{+0.76}$~MeV~\cite{Karliner:2017gml}, while the corresponding value calculated from experimental results is around $100$~MeV.  On the other hand, there is a multitude of the theoretical determinations using various methods such as relativistic quark model~\cite{Ebert:2002ig,Lu:2017meb} and effective potential~\cite{Karliner:2014gca}. Interestingly, they all tend to support the LHCb result rather than the SELEX one. On the side of lattice QCD (LQCD), calculations of the masses are performed by many collaborations~\cite{Flynn:2003vz,Liu:2009jc,Alexandrou:2012xk,Briceno:2012wt,Namekawa:2013vu}, whereas only the result in Ref.~\cite{Alexandrou:2012xk} agrees with the SELEX value. Nevertheless, as pointed out in Ref.~\cite{Namekawa:2013vu}, the chiral extrapolation of the lattice data of Ref.~\cite{Alexandrou:2012xk}, especially the datum at $M_\pi=260$~MeV, using the next-to-leading-order (NLO) heavy baryon chiral perturbation theory would lead to a sizeable systematic uncertainty of the baryon mass in physical limit.  Hence a more appropriate and higher-order extrapolating formula for the masses is required. To that end, we will calculate the masses of the doubly charmed baryons up to next-to-next-to-next-to-leading (N$^3$LO) within the framework of covariant baryon chiral perturbation theory (BChPT).

Chiral perturbation theory (ChPT)~\cite{Weinberg:1978kz,Gasser:1983yg,Gasser:1984gg,Bernard:1995dp} nowadays plays a prominent role in the study of modern hadronic physics at low energies. It has been intensively applied to calculate a multitude of physical quantities and extrapolate lattice QCD data to physical point, see e.g., Ref.~\cite{Yao:2017fym}. Moreover, within ChPT, the finite volume corrections (FVCs) can be systematically obtained by discretizing the integrations involved in the loop contributions~\cite{Leutwyler:1987ak,Gasser:1987zq,Colangelo:2003hf,Beane:2004tw,Beane:2004rf}. For baryon masses, calculations can be performed by using various subtraction methods such as heavy baryon (HB) approach~\cite{Jenkins:1990jv,Bernard:1992qa}, infrared regularization (IR)~\cite{Ellis:1997kc,Becher:1999he} and extended-on-mass-shell (EOMS) scheme~\cite{Gegelia:1999gf,Gegelia:1999qt,Fuchs:2003qc}. Such methods are proposed to settle the power counting issue caused by the presence of non-vanishing baryon mass in the chiral limit, see Refs~\cite{Bernard:2007zu,Geng:2013xn} for reviews.  Nevertheless, the EOMS scheme is more appropriate for the extrapolation of LQCD data. This is because it respects the proper analytical properties when the pion mass is set to certain unphysically large values~\cite{Pascalutsa:2004ga,Pascalutsa:2011fp} and, on the other hand, it leads to results of faster chiral convergence, see, e.g., Refs.~\cite{Fuchs:2003ir,Geng:2008mf,Chen:2012nx,Yao:2016vbz}.  

Within EOMS scheme, though the masses of light baryons have been abundantly studied up to one-loop order, for instance, in Refs.~\cite{Lehnhart:2004vi,MartinCamalich:2010fp,Geng:2011wq,Ren:2012aj,Alvarez-Ruso:2013fza}, the ones of charmed baryons are less investigated. For the doubly charmed baryons, a first calculation of their masses in BChPT was done up to N$^3$LO in Ref.~\cite{Sun:2014aya} using HB formalism. The calculation using EOMS is given in Ref.~\cite{Sun:2016wzh} but only up to next-to-next-to-leading order (N$^2$LO). In the present work,  we extend the calculation up to N$^3$LO. We show explicitly the ultraviolet (UV) divergent and the power counting breaking (PCB) pieces can be absorbed in the low energy constants (LECs). After renormalization, we obtain very compact forms for the mass formulae, which respect correct power counting and also keep proper analytical properties. On top of that, we  derive the relevant FVCs by discretizing the loop contributions. Compared to the results in previous literature, here the so-obtained mass formulae are better suited for chiral extrapolation of LQCD data, especially when more data appears for unphysical values of pion masses. In addition, by imposing the Hellmann-Feynman theorem to the obtained mass formulae, we get the expressions for pion-baryon and strangeness-baryon sigma terms, denoted by $\sigma_{\pi B}$ and $\sigma_{s B}$ with $B\in\{\Xi_{cc},\Omega_{cc}\}$.

As an application, we confront the chiral results of the masses (including FVCs) with the LQCD data of Ref.~\cite{Alexandrou:2012xk} as already mentioned above.  Unfortunately, those data are not sufficient to pin down all the LECs in the
N$^3$LO expressions of masses. Thus we prefer to carry out the numerical analysis with the help of N$^2$LO formulae, where the involved parameters can be well determined through a fit to the data with $M_\pi\leq 500$~MeV. We extrapolate the masses of $\Xi_{cc}$ and $\Omega_{cc}$ to the physical limit and compare them with the existing experimental values. It is found that our result of the $\Xi_{cc}$ mass is in good agreement with the recent determination by LHCb collaboration~\cite{Aaij:2017ueg} within uncertainties. However, it is larger than the value by SELEX collaboration~\cite{Ocherashvili:2004hi}. We predict the sigma terms, $\sigma_{\pi B}$ and $\sigma_{s B}$, as well as the mass splitting between $\Xi_{cc}$ and $\Omega_{cc}$. Their quark mass dependences are also shown for later use when relevant lattice results are available.
 
 This paper is organized as follows. The details of our calculation of the masses and sigma terms within BChPT are elaborated in section~\ref{sec:calc}. In section~\ref{sec:Lag}, the relevant effective Lagrangians are introduced. Chiral results of self-energies and masses together with sigma terms are specified in sections~\ref{sec:se} and \ref{sec:mass}, respectively. Finite volume corrections to the masses are calculated in section~\ref{sec:FVC}.  In section~\ref{sec:num} the numerical study is described. The properties of finite volume corrections are discussed in section~\ref{sec:FVC.properties}.  Fit to lattice QCD data is explained in section~\ref{sec:fit}. In section~\ref{sec:predictions} the prediction of the masses, sigma terms and mass splitting are discussed. Summary is given in section~\ref{sec:sum}. Definition of loop integrals and $\beta$ functions are relegated to Appendices~\ref{sec:loopfuncs} and \ref{sec:betas}, respectively.

\section{Masses and sigma terms in BChPT\label{sec:calc}}
\subsection{Chiral effective Lagrangian\label{sec:Lag}}
The chiral effective Lagrangian relevant for our calculation of the masses and sigma terms up to $\mathcal{O}(p^4)$ can be written as
\bea
\mathcal{L}_{\rm eff}=\mathcal{L}^{(1)}_{\pi\Psi}+\mathcal{L}^{(2)}_{\pi\Psi}+\mathcal{L}^{(4)}_{\pi\Psi}\ ,
\eea
where the numbers in the superscripts denote the chiral orders. 
The leading order (LO) chiral Lagrangian reads
\bea
{\cal L}^{(1)}_{\pi\Psi}= \bar{\Psi}\bigg[i{D}_\mu\gamma^\mu-m+\frac{g_A}{2}{u}_\mu\gamma^\mu\gamma_5\bigg]\Psi\ ,
\eea
where $g_A$ and $m$ are the axial coupling and the mass of the doubly charmed baryons in the chiral limit, respectively.  
According to SU(3) symmetry of light quarks, the doubly charmed baryons of spin-$\frac{1}{2}$ are compiled in the triplet 
\bea
\Psi=
\left(
 \Xi^{++}_{\rm cc} ,\, \Xi^{+}_{\rm cc} ,\, \Omega^{+}_{\rm cc} 
\right)^T \ .
\eea
The covariant derivative acting on the baryon fields is defined by
\bea
D_\mu &=& \partial_\mu+\frac{1}{2}(u^\dagger\partial_\mu u+u\partial_\mu u^\dagger)\ , \nonumber\\
 u&=&{\rm exp}\Big(i\frac{\lambda^a\phi^a}{\sqrt{2}F_0}\Big)\ ,
\eea
where $F_0$ is the decay constant of the Goldstone bosons (GBs) in the chiral limit. The GBs  are collected in the octet
\bea
\lambda^a\phi^a=
\left(
\begin{array}{ccc}
 \frac{1}{\sqrt{2}}\pi^0+\frac{1}{\sqrt{6}}\eta &\pi^+&K^+ \\ 
\pi^- & -\frac{1}{\sqrt{2}}\pi^0+\frac{1}{\sqrt{6}}\eta&K^0\\  
K^-&\bar{K}^0&-\frac{2}{\sqrt{6}}\eta
\end{array}
\right) \ .
\eea
Here the $\lambda^{a}$ ($a=1,\cdots,8$) denote the Gell-Mann matrices and summation over repeated indices is implied. Furthermore, the so-called chiral vielbein $u_\mu$ is given by
\bea
u_\mu=i(u^\dagger\partial_\mu u-u\partial_\mu u^\dagger) \ .
\eea

Analogous to the procedure in Ref.~\cite{Fettes:2000gb}, the NLO Lagrangian is constructed in Ref.~\cite{Sun:2014aya} and has the form
\bea
{\cal L}^{(2)}_{\pi\Psi}&=& \bar{\Psi}\bigg[c_1\langle\chi_+\rangle-\big(\frac{c_2}{8m^2}\langle u_\mu u_\nu\rangle\{D^\mu,D^\nu\}+h.c.\big)\nonumber\\
&-&\big(\frac{c_3}{8m^2}\{ u_\mu, u_\nu\}\{D^\mu,D^\nu\}+h.c.\big)+\frac{c_4}{2}\langle u^2\rangle\nonumber\\
&+&\frac{c_5}{2}u^2+\frac{i\,c_6}{4}\sigma^{\mu\nu}[u_\mu,u_\nu]+c_7\, \hat{\chi}_+
\bigg]\Psi\ .
\eea
Here $\langle\cdots\rangle$ denotes the trace in the flavour space. The chiral block $\chi_+$ is given by
\bea
\chi_+=u^\dagger\chi u^\dagger+u\chi^\dagger u
\eea
with the mass matrix $
\chi=
{\rm diag}(
M_\pi^2 ,  M_\pi^2,2M_K^2-M_\pi^2)
$. The corresponding traceless chiral operator $\hat{\chi}_+$ is defined as $\hat{\chi}_+={\chi}_+-\frac{1}{3}\langle{\chi}_+\rangle$.
The low-energy constants $c_i$ ($i=1,\cdots,7$) are unknown parameters and have dimension GeV$^{-1}$.

At $\mathcal{O}(p^4)$, the following counter terms are needed,
\bea
{\cal L}^{(4)}_{\pi\Psi}= \bar{\Psi}\bigg[e_1\langle\chi_+\rangle^2+e_2\hat{\chi}_+\langle\chi_+\rangle+e_3\langle\hat{\chi}_+^2\rangle+e_4\hat{\chi}_+^2\bigg]\Psi,
\eea
where $e_i$ ($i=1,\cdots,4$) are unknown LECs with mass dimension GeV$^{-3}$.
\subsection{Self-energies of doubly charmed baryons\label{sec:se}}
\begin{figure}[t]
\vspace{1.cm}
\begin{center}
\epsfig{file=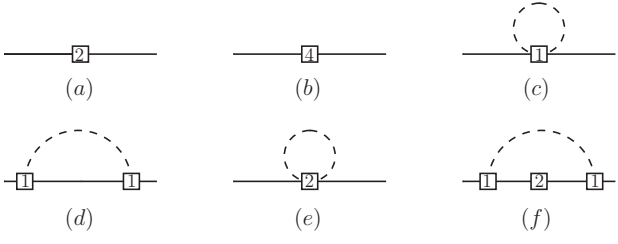,scale=0.6}
\caption{One-particle-irreducible diagrams. Dashed and solid
lines represent pions and nucleons, respectively. Numbers in the squares mark the chiral orders of the vertices. \label{fig:1PI}}
\end{center}
\end{figure}

The one-particle irreducible Feynman diagrams contributing to the baryon two-point functions up to $\mathcal{O}(p^4)$ are displayed in Fig.~\ref{fig:1PI}. 

At $\mathcal{O}(p^2)$, the tree level contribution corresponding to diagram (a) reads
\bea
\Sigma_a^{(2)}(\slashed{p})=-2\big(\hat{c}_1\langle \chi\rangle+c_7 \chi\big)\ ,
\eea
with the combination $\hat{c}_1=c_1-\frac{1}{3}c_7$. The tree contribution of $\mathcal{O}(p^4)$ is from diagram (b) and its explicit expression is
\bea
\Sigma_b^{(4)}(\slashed{p})=-4\big(\hat{e}_1\langle \chi\rangle^2+\hat{e}_2\chi\langle \chi\rangle+e_3\langle \chi^2\rangle+e_4\chi^2\big)
\eea
with $\hat{e}_1=e_1-\frac{1}{3}e_2-\frac{1}{3}e_3+\frac{1}{9}e_4$ and $\hat{e}_2=e_2-\frac{2}{3}e_4$.

At $\mathcal{O}(p^3)$, the leading one-loop order, diagram (c) gives zero contribution, i.e.
$\Sigma_c^{(3)}(\slashed{p})=0$, while diagram (d) yields
\bea
\Sigma_d^{(3)}(\slashed{p})=-\frac{g_A^2}{4F^2}\lambda_a\lambda_a\, { G}_D(\slashed{p},M_a,m)\ ,
\eea
where summation over repeated indices is implied. The loop function $G_D$, together with $G_{E_i,F}$, appearing below, are defined in appendix~\ref{sec:loopfuncs}.

At $\mathcal{O}(p^4)$, the N$^3$LO loop contributions to the self-energy are 
\bea
\Sigma_e^{(4)}(\slashed{p})&=&-\frac{1}{4F^2}\big[4\hat{c}_1\langle\chi\lambda_a\lambda_a\rangle\nonumber\\
&+&c_7\sum_{m=0}^2C_2^m\lambda_a^m\chi\lambda_a^{2-m}\big]{G}_{E1}(\slashed{p},M_a,m)\nonumber\\
&+&\frac{1}{2m^2F^2}\big[c_2\langle\lambda_a\lambda_a\rangle+2c_3\lambda_a\lambda_a\big]{G}_{E2}(\slashed{p},M_a,m)\nonumber\\
&+&\frac{1}{2F^2}\big[c_4\langle\lambda_a\lambda_a\rangle+c_5\lambda_a\lambda_a\big]{G}_{E3}(\slashed{p},M_a,m),
\eea
and 
\bea
\Sigma_f^{(4)}(\slashed{p})=-\frac{g_A^2}{2F^2}\big[\hat{c}_1\lambda_a\langle\chi\rangle\lambda_a+c_7\lambda_a\chi\lambda_a\big]{G}_{F}(\slashed{p},M_a,m)\ .\nonumber \\
\eea

The above self-energies are expressed in matrix form. For a specific doubly charmed baryon $B\in\{\Xi^{++}_{cc},\Xi^{+}_{cc},\Omega^{+}_{cc}\}$ the expression can be obtained using
\bea
\Sigma_B(\slashed{p}) =\chi^T_B\bigg[ \Sigma_a^{(2)}+\Sigma_b^{(4)}+\Sigma_d^{(3)}+\Sigma_e^{(4)}+\Sigma_f^{(4)}\bigg]\chi_B \, ,
\eea
where the unit vectors in the $SU(3)$ flavour space are
\bea
\chi_{\Xi^{++}_{cc}}=
\left(
\begin{array}{c}
 1  \\0\\ 0
\end{array}
\right)
,\, \chi_{\Xi^{+}_{cc}}=\left(
\begin{array}{c}
  0 \\1\\0
\end{array}
\right),\,
\chi_{\Omega^{+}_{cc}}=
\left(
\begin{array}{c}
 0 \\ 0\\1
\end{array}
\right).
\eea

\subsection{The mass and the sigma term \label{sec:mass}}
The dressed propagator $i\,S_B$ of the doubly charmed baryon is expressed as
\bea
iS_B&=&\frac{i}{\slashed{p}-m-\Sigma_B(\slashed{p})}\nonumber\\
&=&\frac{i}{\slashed{p}-m-[\Sigma_B({m})+(
\slashed{p}-m)\Sigma_B^\prime({m})+\mathcal{R}_B(\slashed{p})]}\nonumber\\
&=&\frac{i\,\mathcal{Z}_B}{\slashed{p}-m-\mathcal{Z}_B \Sigma_B({m})-\mathcal{Z}_B\mathcal{R}_B(\slashed{p})}\ ,
\eea
with the wave function renormalization constant
\bea
\mathcal{Z}_B=\frac{1}{1-\Sigma_B^\prime({m})}=1+\Sigma_B^\prime({m})+ {\cal O}(p^4)\,.
\eea
The mass is defined as the pole at $\slashed{p}=m_B$,
\bea
m_B=m+\mathcal{Z}_B\, \Sigma_B({m})+\mathcal{Z}_B\,\mathcal{R}_B(m_B)
\eea
Using the self-energies calculated in the above section and truncating at $\mathcal{O}(p^4)$, one has
\bea\label{eq:massold}
m_B=m+\Sigma_B(m)+\chi^T_B\bigg[\Sigma_a^{(2)}(m) {\Sigma^{(3)}_d}^\prime(m)\bigg]\chi_B
\eea
where the derivative is defined by
\bea
{\Sigma_d^{(3)}}^\prime(\slashed{p})
\equiv-\frac{g_A^2}{4F^2}\lambda_a\lambda_a\, \frac{\partial}{\partial \slashed{p}} {G}_D(\slashed{p},M_a,m)\ .\nonumber
\eea
In Eq.~\eqref{eq:massold}, the UV divergences from loop contributions are subtracted using the modified minimal subtraction ($\widetilde{\rm MS}$) scheme~\cite{Gasser:1987rb} and cancelled by the counter terms generated by the effective Lagrangian. Further, the finite PCB terms due to presence of the internal baryon propagators are absorbed in the LECs.  To that end, one needs to perform the following substitutions of the LECs:
\bea\label{eq:beta}
X&\to& X+\frac{\beta_X\,m\,R}{16\pi^2F^2}+\frac{\bar{\beta}_X\,m}{16\pi^2F^2}\ ,\quad X\in\{m, \hat{c}_1, c_7\}\ ,\nonumber\\
Y&\to& Y+\frac{\beta_Y\,R}{16\pi^2F^2}\ ,\quad Y\in\{\hat{e}_1,\hat{e}_2,e_3,e_4\}\ ,
\eea
where the $\beta$-functions are all given in appendix~\ref{sec:betas}. Here $R={2}/{(d-4)}+\gamma_E-1-\ln(4\pi)$, with $d$ the number of space-time dimensions and $\gamma_E$ the Euler constant.

To be specific, one can organize the explicit expressions of the masses up to $\mathcal{O}(p^4)$ as
\bea\label{eq:massp4}
m_B=m+m_B^{(2)}+m_B^{(3)}+m_B^{(4)}\ ,
\eea
where the $\mathcal{O}(p^2)$ contribution reads
\bea\label{eq:coe2}
m^{(2)}_B=\sum_{\phi=\pi,K}\mathcal{C}_{B,\phi}^{(a)}\, M_\phi^2 \ ,
\eea
with the coefficients $\mathcal{C}_{B,\phi}^{(a)}$ given in table~\ref{tab:coes}. The N$^2$LO corrections to the masses of doubly charmed baryons are
\bea\label{eq:coe3}
m^{(3)}_B=-\sum_{\phi=\pi,K,\eta}\frac{g_A^2}{64\pi^2F^2_{\phi}}\mathcal{C}_{B,\phi}^{(d)}\, {\cal H}_D(M_\phi)\ ,
\eea
while the N$^3$LO ones read
\bea\label{eq:coe4}
m^{(4)}_B&=&\mathcal{C}_{B,\pi}^{(b)}M_\pi^4+\mathcal{C}_{B,K}^{(b)}M_K^4+\mathcal{C}_{B,\pi K}^{(b)}M_\pi^2M_K^2\nonumber\\
&&-\sum_{\phi=\pi,K,\eta}\frac{\mathcal{C}_{B,\phi}^{(e1)}}{64\pi^2F^2_{\phi}}\, {\cal H}_{E1}(M_\phi)\nonumber\\
&&+\sum_{\phi=\pi,K,\eta}\frac{\mathcal{C}_{B,\phi}^{(e2)}}{32\pi^2m^2F^2_{\phi}}\, {\cal H}_{E2}(M_\phi)\nonumber\\
&&+\sum_{\phi=\pi,K,\eta}\frac{\mathcal{C}_{B,\phi}^{(e3)}}{32\pi^2F^2_{\phi}}\, {\cal H}_{E3}(M_\phi)\nonumber\\
&&-\sum_{\phi=\pi,K,\eta}\frac{g_A^2\,\mathcal{C}_{B,\phi}^{(f)}}{32\pi^2F^2_{\phi}}\, {\cal H}_{F}(M_\phi)\nonumber\\
&&-\sum_{\phi=\pi,K,\eta}\frac{g_A^2\,\mathcal{C}_{B,\phi}^{(wf)}}{32\pi^2F^2_{\phi}}\, {\cal H}_{wf}(M_\phi)\ .\label{eq:p4mass}
\eea
All the relevant coefficients are listed in table~\ref{tab:coes}. In appendix~\ref{sec:loopfuncs}, the expressions of the subtracted loop integrals are shown.

 \begin{table}[t]
\caption{\label{tab:coes}Coefficients in the mass formulae: Eqs.~\eqref{eq:coe2}, \eqref{eq:coe3} and \eqref{eq:coe4}. In the table, ${\sigma}_{17}=\hco+c_7$ and  ${\delta}_{17}=\hco-c_7$.\label{tab:coes}}
\vspace{-0.5cm}
\bea
\begin{array}{c|cc}
\hline\hline
&\Xi_{cc}^{++}~(\Xi_{cc}^{+}) & \Omega_{cc}^{+} \\
\hline
\mathcal{C}_{B,\pi}^{(a)}&-2\pos&-2\mos\\
\mathcal{C}_{B,K}^{(a)}&-4\hat{c}_1&-4\pos\\
\hline
\mathcal{C}_{B,\pi}^{(b)}&-4 (\hat{e}_1 + \hat{e}_2 + 3 e_3 + e_4)&-4 (\hat{e}_1 - \hat{e}_2 + 3 e_3 + e_4)\\
\mathcal{C}_{B,K}^{(b)}&-16 (\hat{e}_1 + e_3)&-16 (\hat{e}_1 + \hat{e}_2 + e_3 + e_4)\\
\mathcal{C}_{B,\pi K}^{(b)}&-8 (2 \hat{e}_1 + \hat{e}_2 - 2 e_3)&16 (-\hat{e}_1 + e_3 + e_4)\\
\hline
\mathcal{C}_{B,\pi}^{(d)}&3&0\\
\mathcal{C}_{B,K}^{(d)}&2&4\\
\mathcal{C}_{B,\eta}^{(d)}&\frac{1}{3}&\frac{4}{3}\\
\hline
\mathcal{C}_{B,\pi}^{(e1)}&12(2\hat{c}_1+c_7)M_\pi^2&24\hat{c}_1M_\pi^2\\
\mathcal{C}_{B,K}^{(e1)}&8(4\hat{c}_1+c_7)M_K^2&16(2\hat{c}_1+c_7)M_k^2\\
\mathcal{C}_{B,\eta}^{(e1)}&\frac{4}{3}[6\hco M_\eta^2+c_7M_\pi^2]&\frac{8}{3}[3\pos M_\eta^2-c_7M_\pi^2]\\
\hline
\mathcal{C}_{B,\pi}^{(e2)}&6(c_2+c_3)&6c_2\\
\mathcal{C}_{B,K}^{(e2)}&4(2c_2+c_3)&8(c_2+c_3)\\
\mathcal{C}_{B,\eta}^{(e2)}&\frac{2}{3}(3c_2+c_3)&\frac{2}{3}(3c_2+4c_3)\\
\hline
\mathcal{C}_{B,\pi}^{(e3)}&3(2c_4+c_5)&6c_4\\
\mathcal{C}_{B,K}^{(e3)}&2(4c_4+c_5)&4(2c_4+c_5)\\
\mathcal{C}_{B,\eta}^{(e3)}&\frac{1}{3}(6c_4+c_5)&\frac{2}{3}(3c_4+2c_5)\\
\hline
\mathcal{C}_{B,\pi}^{(f)}&6 \hco M_K^2+3 \pos M_\pi^2&0\\
\mathcal{C}_{B,K}^{(f)}&4 \pos M_K^2 + 2 \mos M_\pi^2&8 \hco M_K^2 + 4 \pos M_\pi^2\\
\mathcal{C}_{B,\eta}^{(f)}&\frac{1}{3}(2 \hco M_K^2 + \pos M_\pi^2)&\frac{4}{3}(2 \pos M_K^2 + \mos M_\pi^2)\\
\hline
\mathcal{C}_{B,\pi}^{(wf)}&-6 \hco M_K^2 - 3 \pos M_\pi^2&0\\
\mathcal{C}_{B,K}^{(wf)}&-4 \hco M_K^2 - 2 \pos M_\pi^2&-8 \pos M_K^2 - 4 \mos M_\pi^2\\
\mathcal{C}_{B,\eta}^{(wf)}&-\frac{1}{3}[2 \hco M_K^2 + \pos M_\pi^2]&-\frac{4}{3} [2 \pos M_K^2 + \mos M_\pi^2]\\
\hline\hline
\end{array}\nonumber
\eea
\end{table}

The sigma terms can be obtained by applying the Hellmann-Feynman theorem to the masses,
\bea
\sigma_{\pi B}=\hat{m}\frac{\partial m_B}{\partial \hat{m}}\ ,\quad
\sigma_{sB}=m_s\frac{\partial m_B}{\partial m_s}\ ,
\eea
where  $\hat{m}=(m_u+m_d)/2$.  Here the up, down and strange quark masses are denoted by $m_{u}$, $m_d$ and $m_s$, respectively. In the isospin limit,  i.e. $m_u=m_d=\hat{m}$, the quark masses are simply related to the LO masses of the GBs though: 
\bea\label{eq:LOGBmass}
M_\pi^2&=&2B_0\hat{m}\ ,\quad M_K^2=B_0(\hat{m}+m_s)\ ,\nonumber\\
M_\eta^2&=&2B_0(\hat{m}+2m_s)/3 \ ,
\eea
with $B_0$ being a constant related to quark condensate. Therefore, in practice, the derivatives can be rewritten with respect to the GBs masses, instead of the quark masses.

\begin{figure*}[t]
\vspace{1.cm}
\begin{center}
\epsfig{file=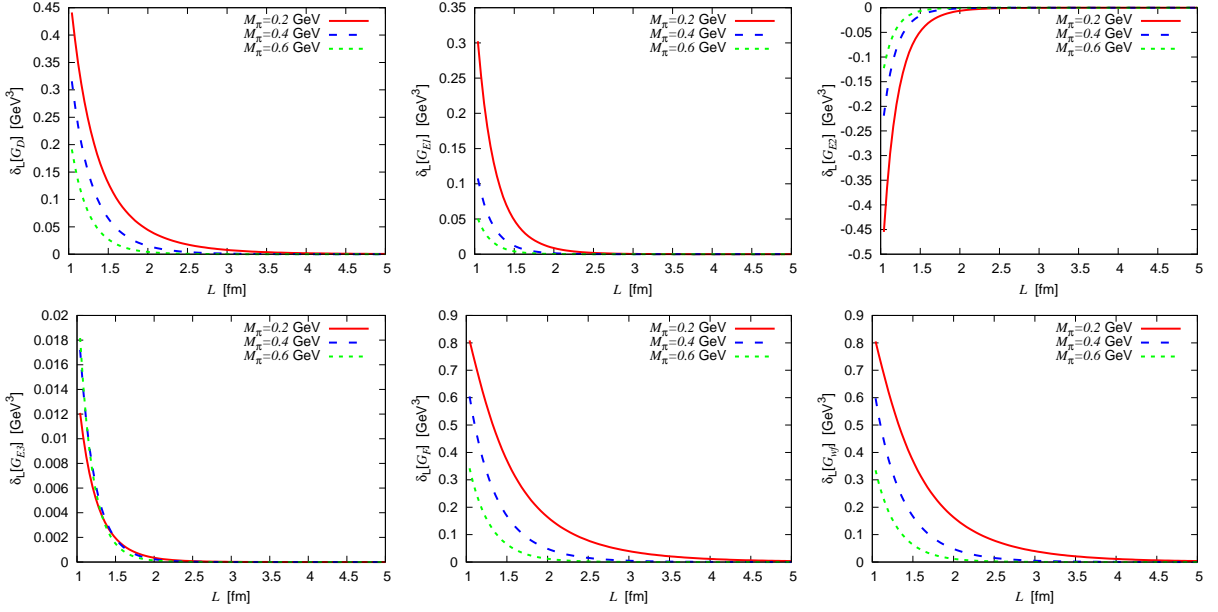,scale=0.75}
\caption{Finite volume corrections to loop integrals as functions of the lattice size $L$.\label{fig:FVC}}
\end{center}
\end{figure*}

\begin{figure*}[t]
\vspace{1.cm}
\begin{center}
\epsfig{file=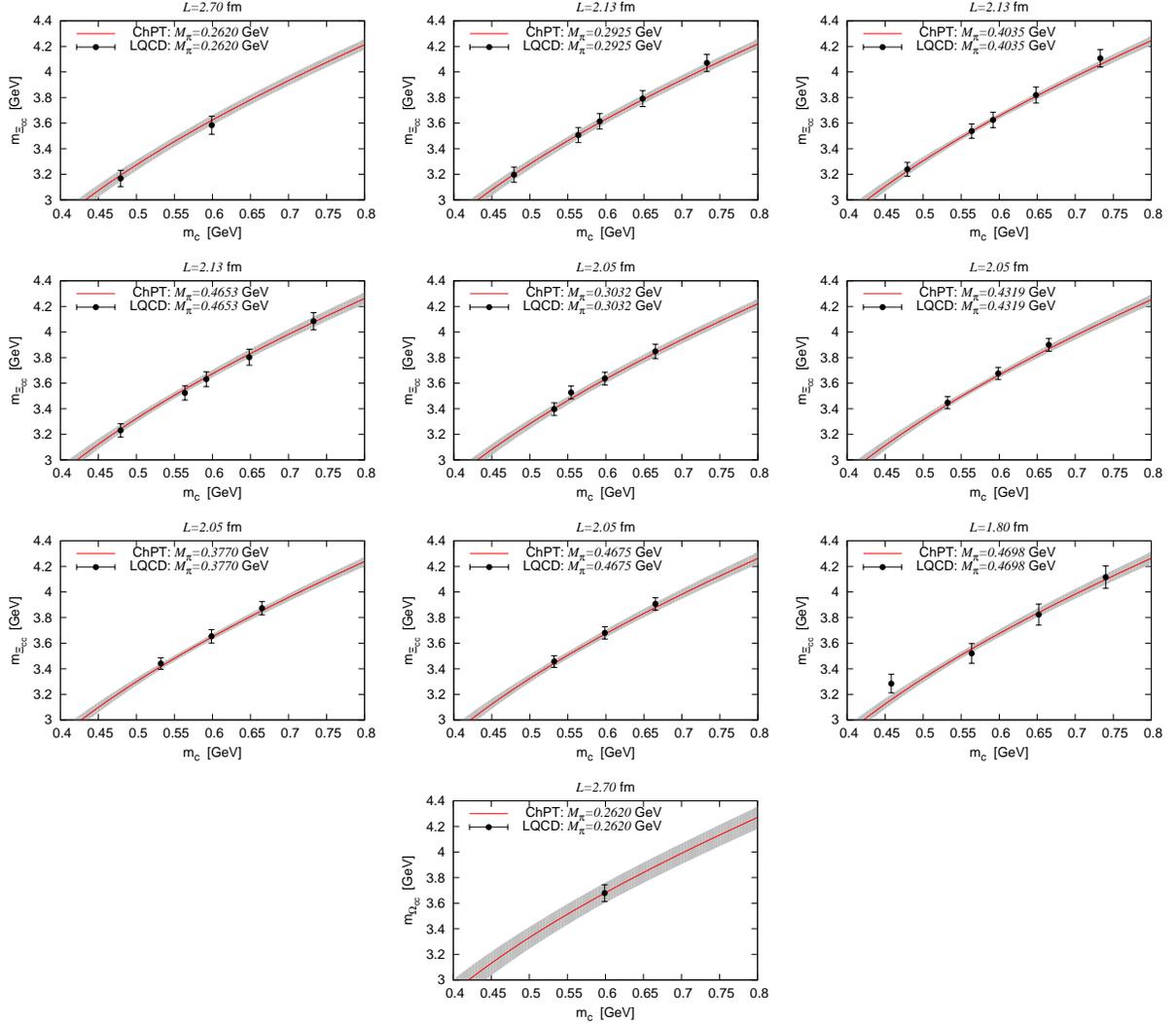,width=0.9\textwidth}
\caption{\label{fig:latmas}Masses of doubly charmed baryons as functions of $m_c$ for different pion masses and lattice sizes.}
\end{center}
\end{figure*}

\begin{figure}[t]
\vspace{1.cm}
\begin{center}
\epsfig{file=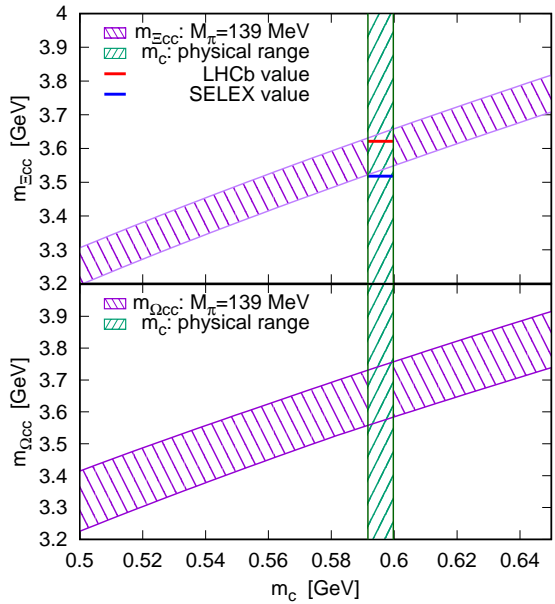,scale=1.1}
\caption{\label{fig:dcmassphy}Physical masses of the doubly charmed baryons.}
\end{center}
\end{figure}

\begin{figure*}[t]
\vspace{1.cm}
\begin{center}
\epsfig{file=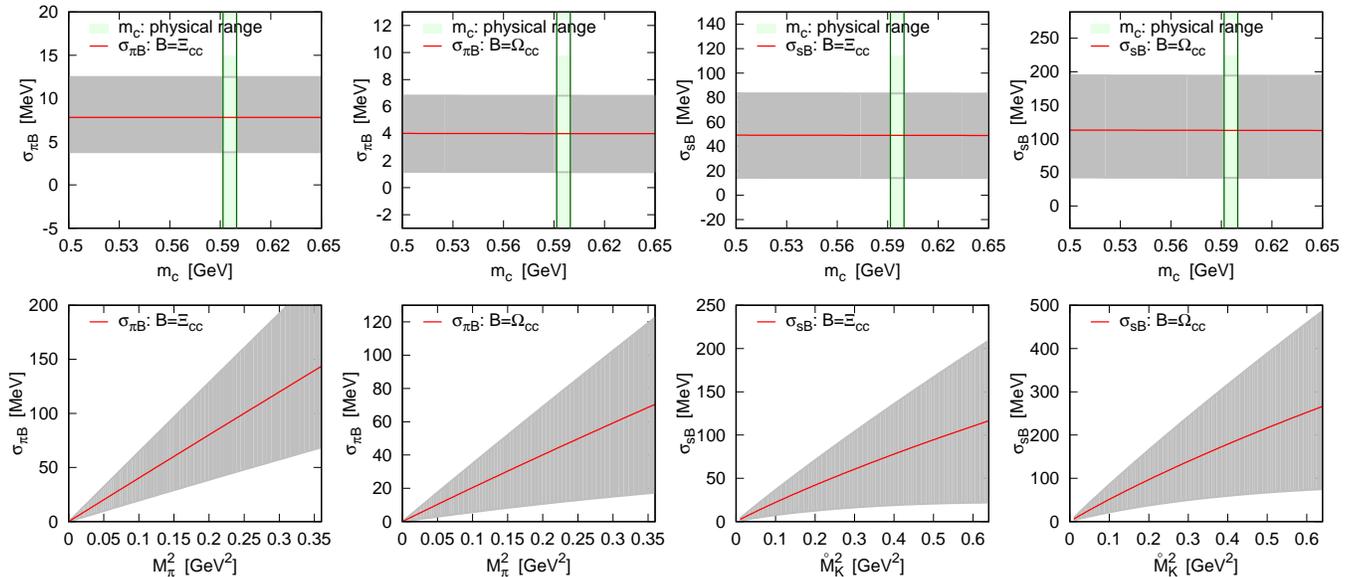,width=\textwidth}
\caption{\label{fig:sigma}Quark mass dependences of the sigma terms for the doubly charmed baryons.}
\end{center}
\end{figure*}

\begin{figure}[t]
\vspace{1.cm}
\begin{center}
\epsfig{file=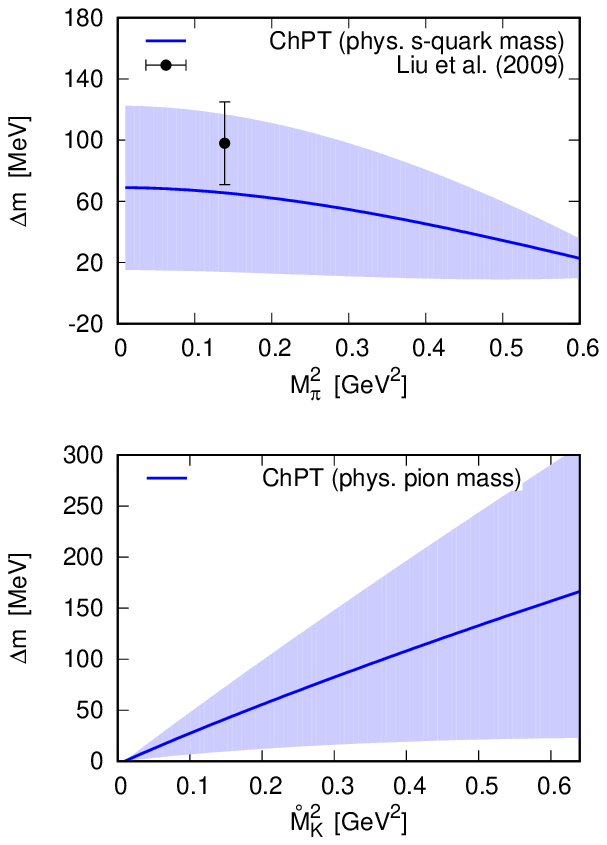,scale=1.2}
\caption{\label{fig:spliting}Quark mass dependence of the mass splitting. }
\end{center}
\end{figure}

\subsection{Finite volume corrections\label{sec:FVC}}

On the lattice, simulations are performed for a system of interest enclosed in a finite box. The momentum is discretized and can only take values of $2\pi\vec{n}/L$ with $\vec{n}$ a vector of integers and $L$ the side length of the hypercube. Consequently, an integration over spatial momenta in infinite volume corresponds to a summation over the momentum modes in finite volume. The difference caused by such a replacement is named as finite volume correction. Specifically, the finite volume correction for a given quantity $\mathcal{Q}$ is given by 
\bea
\delta_L[\mathcal{Q}]=\mathcal{Q}(L)-\mathcal{Q}(\infty)\ ,
\eea 
where $\mathcal{Q}(L)$ and $\mathcal{Q}(\infty)$ are calculated in finite volume $L^3$ and infinite volume, respectively. In the so-call $p$-regime where $M_\phi L\gg 1$, ChPT provides a systematical tool to investigate finite-volume dependence of observables. To that end, one just needs to calculate the integrals stemming from loop diagrams in a finite box, while the temporal dimension can be treated as infinite since it is generally much larger than the spatial components in LQCD simulation for zero-temperature. 

To obtain finite volume corrections to the masses of doubly charmed baryons, we choose to work in the rest frame of the baryons and follow the procedure demonstrated in Ref.~\cite{Beane:2004tw,Geng:2011wq}. For the loop integral $\mathcal{H}_D$, we obtain
\bea\label{eq:deltaLHD}
\delta_L[\mathcal{H}_{D}]&=&\int_0^1{\rm d}x\bigg\{m{\left(\frac{1}{2}+x\right)}\delta_{\frac{1}{2}}(L,{\cal M}^2_B)\\
&-&\frac{m}{4}\big[x^3m^2+(2+x){\cal M}^2_B\big]\delta_{\frac{3}{2}}(L,{\cal M}^2_B)\bigg\}\ ,\nonumber
\eea
with $\mathcal{M}_B^2=x^2 m^2+(1-x)M_\phi^2-i0^+$. Here the integration is performed over the Feynman parameter $x$. Furthermore, the master function is given by
\bea
\delta_{r}(L,\mathcal{M}^2)&=&\frac{2^{-1/2-r}(\sqrt{\MM})^{3-2r}}{\pi^{3/2}\Gamma(r)}\sum_{n=1}^{\infty}{\rm Mul}(n)\nonumber\\
&\times&\big(L\sqrt{\MM}\sqrt{n}\big)\,\mathcal{K}_{3/2-r}(L\sqrt{\MM}\sqrt{n})\,,
\eea
where $\mathcal{K}_r(z)$ is the modified Bessel function of the second kind, and ${\rm Mul}(n)$ is multiplicity whose value up to $n=20$ can be found in, e.g., Ref.~\cite{Colangelo:2003hf}. Analogously, the FVCs for $\mathcal{H}_{Ei}$ ($i=1,2,3$) read
\bea
\delta_L[\mathcal{H}_{E1}]&=&-\frac{1}{2}\delta_{\frac{1}{2}}(L,{M}^2_\phi)\ ,\\
\delta_L[\mathcal{H}_{E2}]&=&-\frac{1}{2}m^2\delta_{-\frac{1}{2}}(L,{M}^2_\phi)\ ,\\
\delta_L[\mathcal{H}_{E3}]&=&-\frac{1}{2}M_\phi^2\delta_{\frac{1}{2}}(L,{M}^2_\phi) \ .
\eea
There are no integrations over Feynman parameters in the above expressions since only one internal propagator is involved in each tadpole loop. The calculation of the FVCs corresponding to diagram (f) in Fig.~\ref{fig:1PI} is more complicated because of the presence of three internal propagators. Nonetheless, the result can be obtained straightforwardly, which is
\bea
\delta_L[\mathcal{H}_{F}]&=&-\int_0^1{\rm d}x\,\, x\bigg\{\delta_{\frac{1}{2}}(L,{\cal M}^2_B)-\frac{1}{2}\big[3m^2(1+x^2)\nonumber\\
&+&2{\cal M}^2_B\big]\delta_{\frac{3}{2}}(L,{\cal M}^2_B)+\frac{3}{8}\big[x^4m^4+2m^2{\cal M}^2_B(2+x^2)\nonumber\\
&+&{\cal M}^4_B\big]\delta_{\frac{5}{2}}(L,{\cal M}^2_B)\bigg\}\ ,
\eea
where $\mathcal{M}_B$ is the same as the one in Eq.~\eqref{eq:deltaLHD}. Lastly, the contribution due to the wave function renormalization is given by
\bea
\delta_L[\mathcal{H}_{wf}]&=&\int_0^1\frac{{\rm d}x}{4}\bigg\{4x\delta_{\frac{1}{2}}(L,{\cal M}^2_B)-\big[m^2x(9x^2-x-6)\nonumber\\
&+&(1+x){\cal M}^2_B\big]\delta_{\frac{3}{2}}(L,{\cal M}^2_B)+3m^2(x-1)x\nonumber\\
&\times&\big[m^2x^3+(2+x){\cal M}^2_B\big]\delta_{\frac{5}{2}}(L,\mathcal{M}_B^2)\bigg\}\ .
\eea

In the end of this section, it is worth stressing that the calculations of FVCs are performed in four dimensions: a finite hypercube plus an infinite time interval. This is feasible due to the fact that $\mathcal{Q}(L)$ and $\mathcal{Q}(\infty)$ have the same ultraviolet property which guarantees that $\delta_L[\mathcal{Q}]$ is finite in four dimensions. Besides, as pointed out in Ref.~\cite{Geng:2011wq}, there are no PCB terms in $\delta_L[\mathcal{Q}]$ either, since the short-distance behaviours of $\mathcal{Q}(L)$ and $\mathcal{Q}(\infty)$ should be exactly identical. Thus, the quantities respecting power counting in finite volume can be easily obtained just by adding the FVCs to the corresponding EOMS-renormalized ones in infinite volume.

\section{Numerical results and discussion\label{sec:num}}

\subsection{Properties of finite volume corrections\label{sec:FVC.properties}}

We compute the finite volume corrections given in section~\ref{sec:FVC} as functions of the lattice size $L$ with three different Goldstone masses $M_\phi=0.2,\, 0.4$ and $0.6$~GeV. The baryon mass $m$ is fixed to $3.6$~GeV. The results are shown in Fig.~\ref{fig:FVC}. From the figure, on the one hand, it is found that all the relevant FVCs decrease rapidly as $L$ increases up to $\sim 2$~fm, behaving quite typically as the FVCs for the nucleons shown in Ref.~\cite{Geng:2011wq}. The lattice QCD data used in our fit are obtained using lattice spaces ranging from $1.8$~fm to $2.7$~fm, which are in the vicinity of the turning point. The data corresponding to $L=1.8$~fm might receive a larger FVC than the others.  
On the other hand, the smaller the Goldstone mass $M_\phi$ is, the bigger the modules of the FVCs are. Therefore, contributions due to coupling of light pions dominate and for lattice data FVCs are larger when simulations are done with values of masses close to physical ones.  Note that we checked that the influence of changing the baryon mass $m$, e.g., in the range $[2.6,4.6]$~GeV, is negligible.

The finite volume corrections $\delta_L[\mathcal{H}_F]$ and $\delta_L[\mathcal{H}_{wf}]$ are rather similar. Both of them are respectively larger than the other ones in Fig.~\ref{fig:FVC}. Nonetheless, for the FVCs to the masses in Eq.~\eqref{eq:p4mass}, there should exist sizeable cancellation between the two relevant terms in the last two rows, since their corresponding coefficients have opposite signs, as can be seen in table~\ref{tab:coes}.

\subsection{Fit to lattice QCD data\label{sec:fit}}

We are now in the position to confront the chiral expression of doubly charmed baryons with lattice QCD determinations by explicitly including finite volume corrections. As already discussed in the introduction, it is interesting to study the lattice QCD data given in Ref.~\cite{Alexandrou:2012xk}. Unfortunately, in our theoretical formula, Eq.~\eqref{eq:massp4}, there are too many unknown LECs, twelve in total: $m$, $c_i$ ($i=1,\cdots,5,7$), $g_A$ and $e_{j}$ ($j=1,\cdots,4$). Hence, we start with mass formula just at $\mathcal{O}(p^3)$ order where only four parameters, $m$, $c_{1,7}$ and $g_A$, are involved.

The lattice QCD data are obtained by numerical simulations with unphysical quark masses. The $u$-, $d$- and $s$-quark mass dependence can be always expressed in terms of the dependence on the leading-order masses of the Goldstone bosons shown in Eq.~\eqref{eq:LOGBmass}. More specifically, the light $u$- or $d$-quark mass dependence is usually re-expressed as pion mass dependence. The $s$-quark mass dependence can be casted to the kaon mass in the limit of $M_\pi^2(\propto\hat{m})\to0$, denoted as $\mathring{M}_K^2$. Then, with the help of Eq.~\eqref{eq:LOGBmass}, the pion- and strange-mass dependence of the kaon mass can be written as 
\bea
M_K=\sqrt{\mathring{M}_K^2+M_\pi^2/2}\ , \qquad\mathring{M}_K^2=B_0m_s.
\eea
The data for the strange-doubly-charmed baryon $\Omega_{cc}$ is obtained with a strange quark mass very close to the tuned value using physical kaon mass
~\cite{Alexandrou:2012xk}. Therefore, as a good approximation, one can fix $\mathring{M}_K^2$ just by imposing the physical values of the pion and kaon masses: $M_\pi^{\rm phy}=139$~MeV and $M_K^{\rm phy}=496$~MeV.
As for $M_\eta$, it is always obtained from pion and kaon masses through the Gell-Mann-Okubo mass relation: $3M_\eta^2=4M_K^2-M_\pi^2$. The masses of doubly charmed baryons also depend on the valence c-quark mass. In SU(3) chiral limit, all the light quark masses are zero and the baryon mass is equal to $m$, i.e., the first term on the right hand side of Eq.~\eqref{eq:massp4}. It is thus reasonable to assume that only the chiral-limit baryon mass, $m$, carries the information of the dependence on the $c$ quark mass.  In line with heavy quark expansion, such a dependence can be expressed in the form of 
\bea
m=\tilde{m}+2\,m_c+{\alpha}/{m_c}+\mathcal{O}({1}/{m_c^2})\ ,
\eea
where $\tilde{m}$ and $\alpha$ are unknown constants. Since the QCD data of Ref.~\cite{Alexandrou:2012xk} are provided for various values of the $c$-quark mass, those two constants should be treated as fitting parameters, instead of $m$. 

In our fitting procedure, we employ $F_\pi=92.2$~MeV, $F_K=112$~MeV and $F_\eta=110$~MeV as done in Ref.~\cite{Sun:2014aya}. The pion mass dependences of the decay constants are not taken into account, since the caused differences are of higher orders - at least $\mathcal{O}(p^5)$.
Furthermore, the axial coupling constant is fixed to $g_A=-0.2$~\cite{Sun:2016wzh}. The $g_A$ here is related to a common coupling $g$ involved in an effective Lagrangian respecting heavy quark-diquark symmetry~\cite{Hu:2005gf}, whose value can be further estimated by fitting to the $D^{\ast+}$ decay width. Finally, it is better to use the combination $\hat{c}_1$ rather than $c_1$ as a fitting parameter such that possible large correlation between $c_1$ and $c_7$ can be avoided.
In summary, the fitting parameters in our fit at $\mathcal{O}(p^3)$ order are $\tilde{m}$, $\alpha$, $\hat{c}_1$ and $c_7$. 

\begin{table}[ht]
\caption{Fit results (with finite volume corrections). }\label{tab:fitFVC}
\vspace{-0.5cm}
\bea
\begin{array}{cr|cccc}
\hline\hline
 &\text{Value}&\multicolumn{4}{c}{\text{Correlation matrix}}\\
\cline{3-6}
\chi^2{\rm/d.o.f}&\frac{5.29}{35-4}~~~~& \tilde{m} & \alpha  & \hat{c}_1 & c_7 \\
\hline
\tilde{m}~~[{\rm GeV}]&3.101(0.111)&1&-0.68&0.69 & -0.19\\
\alpha~~[{\rm GeV}^2]&-0.453(0.047)&&1&0.01&-0.05\\
\hat{c}_1~~[{\rm GeV}^{-1}]&-0.064(0.055)&&&1&-0.56\\
{c}_7~~[{\rm GeV}^{-1}]&-0.085(0.085)&&&&1\\
\hline\hline
\end{array}\nonumber
\eea
\end{table}

To proceed, we perform fit to the lattice QCD data corresponding to different values of $M_\pi$, $m_c$ and $L$. The best-fitted results of the parameters and their correlations are collected in table~\ref{tab:fitFVC}. As one can see from the table, the values come out to be very natural and the correlations are quite acceptable. It is also found that the inclusion of FVCs does not change the values of the parameters dramatically when compared to the results in Ref.~\cite{Sun:2016wzh} obtained regardless of FVCs. In Fig.~\ref{fig:latmas} we plot the masses of $\Xi_{cc}$ and $\Omega_{cc}$ as functions of $m_c$ for different pion masses $M_\pi$ and lattice sizes $L$. The grey bands are obtained by varying the parameters within their 1-$\sigma$ uncertainties. All the data with $M_\pi\leq 500$~MeV are included in the fit. The fit results remain almost the same if we lessen the range of pion mass to $M_\pi\leq 400$~MeV. It is not feasible to decrease further the range as the data included are not sufficient to achieve a stable fit.

The above discussions are dedicated to the fit using mass formula truncated at $\mathcal{O}(p^3)$. Extension to $\mathcal{O}(p^4)$ is straightforward. Nonetheless, similar to the case for nucleon mass at  $\mathcal{O}(p^4)$~\cite{Ren:2012aj}, one has to replace the LO meson masses in $m_B^{(2)}$ by their corresponding $\mathcal{O}(p^4)$ counterparts, which can be found, for instance, in Ref.~\cite{Gasser:1984gg}. Such a replacement generate $\mathcal{O}(p^4)$ contributions to $m_B^{(4)}$. The relevant LECs of $L_i$ in $\mathcal{O}(p^4)$ Goldstone masses can be fixed to the empirical values given in Ref.~\cite{Bijnens:2014lea}.  We fitted to the lattice QCD data but no stable results can be achieved. The data set is not sufficient to pin down twelve fitting parameters.

\subsection{Predictions\label{sec:predictions}}
We can make predictions based on the fitted values of the parameters in table~\ref{tab:fitFVC}. In Fig.~\ref{fig:dcmassphy}, the masses of the doubly charmed baryons are plotted as functions of $m_c$ with $M_\pi=M_{\pi}^{\rm phy}$ and $L\to\infty$. In Ref.~\cite{Alexandrou:2012xk}, three different values of lattice spacing are used in the simulations, which are denoted by $\beta=3.9$, $\beta=4.05$ and $\beta=4.2$. The corresponding physical values of the charm quark mass are $m_c^{\rm phy}[\beta_1]=0.598$~GeV, $m_c^{\rm phy}[\beta_2]=0.591$~GeV and $m_c^{\rm phy}[\beta_3]=0.598$~GeV, respectively. We take the average as the central value of $m_c^{\rm phy}$ and the standard deviation as the error, which leads to $m_c^{\rm phy}=0.596(4)$.  Correspondingly, in Fig.~\ref{fig:dcmassphy}, the vertical green slashed band corresponds to the physical region of $m_c$ within its $1$-$\sigma$ standard deviation. In addition, the purple back-slashed band is obtained by varying the parameters within their $1$-$\sigma$ uncertainties given in table~\ref{tab:fitFVC}. Our predicted physical masses of the baryons are located in the overlaps of the two bands. In the top panel of Fig.~\ref{fig:dcmassphy}, we also show the experimental values of the mass of $\Xi_{cc}$ by LHCb~\cite{Aaij:2017ueg} and SELEX~\cite{Mattson:2002vu}. Interestingly, it is found that our prediction is in good agreement with the LHCb determination. On the contrary, the SELEX value is just below the border of our predicted region. 
{For easy reference, our predicted physical masses and sigma terms of the doubly charmed baryon are compiled in Table~\ref{tab:mass}. Note that the predicted strangeness sigma terms of the doubly charmed baryon are comparable to those of the ground-state octet baryons, see, e.g., in Ref.~\cite{Ren:2014vea}. For instance, the strangeness sigma terms of the $\Xi_{cc}$ and the nucleon,  i.e., states without valence $s$ quark in quark-model interpretation, turn out to be of the same order, i.e., tens of MeV. However, regarding the pion sigma terms, unlike the case for the nucleon that a large value of $\sigma_{\pi N}$ ( $\ge50$~MeV) was obtained~\cite{Alarcon:2011zs,Alarcon:2012nr,Hoferichter:2015dsa,Ling:2017jyz}, our predicted values of $\sigma_{\pi\Xi_{cc}}$ and $\sigma_{\pi\Omega_{cc}}$ are small. }

\begin{table}[ht]
\caption{Physical masses and sigma terms. }\label{tab:mass}
\vspace{-0.5cm}
\bea
\begin{array}{c|cc}
\hline\hline
&B=\Xi_{cc}&B=\Omega_{cc}\\
\hline
m_B & 3.591\pm 0.067~{\rm GeV} & 3.657\pm0.100~{\rm GeV}\\
\sigma_{\pi B}&10.5\pm 3.4~{\rm MeV}&4.0\pm 2.8~{\rm MeV}\\
\sigma_{s B}&48.7\pm 34.7~{\rm MeV}&118.0\pm 76.1~{\rm MeV}\\
\hline\hline
\end{array}\nonumber
\eea
\end{table}

Likewise, the $m_c$-dependence of sigma terms, at physical pion mass and in infinite volume, are shown in first line of Fig.~\ref{fig:sigma}. The grey bands are due to the variation of the fitted parameters within their uncertainties. The vertical green bands represent the physical $m_c$ region. Our predicted values for the sigma terms are inside the overlaps. Unlike the masses in Fig.~\ref{fig:dcmassphy}, which strongly depend on $m_c$, one can notice from Fig.~\ref{fig:sigma} that the dependence of sigma terms on $m_c$ is negligible. In other words, the influence of the heavy $c$ quark is almost absent for the sigma terms. This observation verifies that the sigma terms are more appropriate than the masses to explore the chiral dynamics of the doubly charmed baryons. In the second line of Fig.~\ref{fig:sigma} 
we show $M_\pi^2(\propto\hat{m})$ dependence of $\sigma_{\pi B}$ where the other quark masses are set to physical values. As expected, the values of $\sigma_{\pi B}$  with $B=\Xi_{cc},\Omega_{cc}$ increase with $M_\pi^2$. We checked also that the $\sigma_{sB}$ are not sensitive to the variation of $M_\pi^2$. Nonetheless, there exists strong $\mathring{M}_K^2(\propto{m}_s)$ dependence for $\sigma_{sB}$ as one can see from the last two plots in Fig.~\ref{fig:sigma}.

Another interesting quantity related to the light quarks is the mass splitting between the $\Xi_{cc}$ and $\Omega_{cc}$. In Fig.~\ref{fig:spliting} the $M_\pi^2$ and $\mathring{M}_K^2$ dependences of the mass splitting $\Delta m$ are shown. It is found that $\Delta m$ depends more strongly on $\mathring{M}_K^2$ than $M_\pi^2$. Furthermore, the different trends of $\Delta m$ as the quark masses increase
 validate the fact that $\Delta m\propto m_s-\hat{m}$. At physical quark masses  our prediction is $\Delta m=65.9\pm51.3$~MeV, in agreement with the determination extrapolated by the Lattice QCD group of Ref~\cite{Liu:2009jc}.

\section{Summary\label{sec:sum}}
We have calculated the masses and sigma terms of the doubly charmed baryons up to $\mathcal{O}(p^4)$ in a covariant baryon chiral perturbation theory with Goldstone bosons and the baryons as degrees of freedom. The masses at complete one-loop order is renormalized by making use of the EOMS scheme, which restores the correct power counting while respecting the proper analytic structure.  As a consequence, we also obtained the pion-baryon and strangeness-baryon sigma terms by applying Hellmann-Feynman theorem to the obtained masses.  In order to make comparison with LQCD results in a more rigorous manner, the finite volume corrections to the chiral results of the masses are derived systematically by discretizing the loop contributions.  FVCs corresponding to the relevant loop integrals are studied numerically and typical behaviour when varying the lattice size $L$ is observed, namely, FVCs decrease rapidly as $L$ increases up to $\sim 2$~fm.

Using the mass formulae with FVCs, we investigated the pion-mass and $m_c$ dependences for the masses of doubly charmed baryons by performing fits to lattice QCD data of Ref.~\cite{Alexandrou:2012xk}. It is found that more data, with respect to more values of unphysical pion masses, are required to pin down the LECs appearing in the N$^3$LO formulae. Nevertheless, the LECs in the N$^2$LO mass expressions can be well determined. Based on the fitted values, we have extrapolated the baryon masses to the physical limit. We find that our result for $m_{\Xi_{cc}}$ is in agreement with the latest experiment determination by LHCb collaboration within uncertainty. However, it is larger than the value by SELEX collaboration. Finally, we predict the sigma terms $\sigma_{\pi B}$ and $\sigma_{s B}$ with $B\in\{\Xi_{cc},\Omega_{cc}\}$, as well as the mass splitting between $\Xi_{cc}$ and $\Omega_{cc}$ states. Their quark mass dependences are studied as well. 

The masses calculated in the present work will be useful in the future investigation of observables like axial charge and scattering lengths, related to the doubly charmed baryons, within the framework of covariant BChPT, since they are basic quantities involved in expressions of almost all the others. The sigma terms are related to the potentials of the GBs scattering off the doubly charmed baryons, and hence can be implemented as an additional constraint when making prediction of exotic doubly charmed baryons based on unitatized potentials.

\acknowledgements
The author would like to thank Xiu-Lei~Ren and Zhi-Feng~Sun for helpful discussions. He is also grateful to Jambul~Gegelia and Manuel~Jose~Vicente-Vacas for reading the manuscript and giving valuable comments. DLY acknowledges the warm hospitality of the ITP of CAS where this paper was finalized. This research is supported by the Spanish Ministerio de Econom\'ia y Competitividad and the European Regional Development Fund, under contracts FIS2014-51948-C2-1-P, FIS2014-51948-C2-2-P, SEV-2014-0398 and by Generalitat Valenciana under contract PROMETEOII/2014/0068.

\appendix
\section{Loop integrals}
\label{sec:loopfuncs}
The loop functions involved in the self-energies in section~\ref{sec:se} are defined as follows:
\bea
G_D(\slashed{p},M_a,m)&\equiv& \frac{1}{i}\int\frac{d^dk}{(2\pi)^d}\frac{\slashed{k}[(\slashed{k}+\slashed{p})-m]\slashed{k}}{[k^2-M_a^2][(k+p)^2-m^2]}\ ,\nonumber\\
{G}_{E1}(\slashed{p},M_a,m)&\equiv&\frac{1}{i}\int\frac{d^dk}{(2\pi)^d}\frac{1}{k^2-M_a^2}\ ,\nonumber\\
{G}_{E2}(\slashed{p},M_a,m)&\equiv&\frac{1}{i}\int\frac{d^dk}{(2\pi)^d}\frac{(k\cdot p)^2}{k^2-M_a^2}\ ,\nonumber\\
{G}_{E3}(\slashed{p},M_a,m)&\equiv&\frac{1}{i}\int\frac{d^dk}{(2\pi)^d}\frac{k^2}{k^2-M_a^2}\ ,\nonumber\\
{ G}_F(\slashed{p},M_a,m)&\equiv& \frac{1}{i}\int\frac{d^dk}{(2\pi)^d}\frac{\slashed{k}[(\slashed{k}+\slashed{p})-m]^2\slashed{k}}{[k^2-M_a^2][(k+p)^2-m^2]^2}\ .\nonumber
\eea

The loop integral in the $\mathcal{O}(p^3)$ mass formula~\eqref{eq:coe3} is
\bea
{\cal H}_D(M_\phi)&=&\frac{2 M_\phi^2}{m} \bigg\{\IB + m^2 \big[1 - \IBP(m^2)\big]\bigg\}\ ,
\eea
while the ones in the $\mathcal{O}(p^4)$ mass formula~\eqref{eq:coe4} read 
\bea
{\cal H}_{E1}(M_\phi)&=&\IP \ ,\nonumber\\
{\cal H}_{E2}(M_\phi)&=&\frac{1}{8}M_\phi^2m^2\big[M_\phi^2+2\IP\big]\ ,\nonumber\\
{\cal H}_{E3}(M_\phi)&=&M_\phi^2\IP\ ,\nonumber\\
{\cal H}_F(M_\phi)&=&\frac{1}{4 m^2 - M_\phi^2}\bigg\{4 M_\phi^2 \IB + ( M_\phi^2-12 m^2 ) \IP \nonumber\\
    &-& 2 M_\phi^2 \big[2 m^2 + (  M_\phi^2-6 m^2) \IBP(m^2)\big]\bigg\}\ ,\nonumber\\
{\cal H}_{wf}(M_\phi)&=&\frac{1}{4 m^2 - M_\phi^2}\bigg\{ ( 5M_\phi^2-12 m^2 ) \IP-4 M_\phi^2 \IB \nonumber\\
    &+& 4 M_\phi^2 \big[m^2 + ( 3m^2- M_\phi^2) \IBP(m^2)\big]\bigg\}\ .
\eea
Above, the one-loop scalar integrals are defined by
\bea
&&\IP=-M_\phi^2\ln\frac{M_\phi^2}{\mu^2} ,\qquad \IB=-m^2\ln\frac{m^2}{\mu^2}\ , \nonumber\\
&&\IBP(p^2)=
1-\ln\frac{m^2}{\mu^2}+
\frac{M_\phi^2-m^2+p^2}{2\,p^2}\ln\frac{m^2}{M_\phi^2} \nonumber\\
&&\hspace{0.7cm}+
\frac{p^2-(M_\phi-m)^2}{p^2}\rho_{\phi}(p^2)
\ln\frac{\rho_{\phi}(p^2)-1}{\rho_{\phi}(p^2)+1}\ ,
\eea
with $\mu$ the renormalization scale and 
\bea
\rho_{\phi}(p^2)\equiv\sqrt{\frac{p^2-(M_\phi+m)^2}{p^2-(M_\phi-m)^2}}.
\eea
In our numerical calculation, $\mu$ is set to $1$~GeV.

\section{$\beta$ functions\label{sec:betas}}
In Eq.~\eqref{eq:beta}, the $\beta$ functions involved in the cancellation of UV divergences are
\bea
\beta_m&=&\frac{8}{3}g_A^2m^2\ ,\nonumber\\
\beta_{\hat{c}_1}&=&-\frac{11}{36}g_A^2+(8\hat{c}_1+3c_7)g_A^2m\ ,\nonumber\\
\beta_{{c}_7}&=&-\frac{5}{12}g_A^2-c_7g_A^2m\ ,\nonumber\\
\beta_{\hat{e}_1}&=&\frac{4 (c_3 - 33 c_4 + 2 c_5) + 264 \hat{c}_1 (1 + g_A^2)-33 c_2 }{864}\ ,\nonumber\\
\beta_{\hat{e}_2}&=&\frac{-13 c_3 - 26 c_5 + 44 c_7 + (15 \hat{c}_1 + 11 c_7)4 g_A^2}{144 }\ ,\nonumber\\
\beta_{e_3}&=&\frac{1}{288}\big[120 \hat{c}_1 - 15 c_2 - 13 c_3 - 60 c_4 \nonumber\\
&&\qquad- 26 c5 + 36 c_7 (1+g_A^2) \big]\ ,\nonumber\\
\beta_{e_4}&=&\frac{3 c_3 + 6 c_5 + 4 c_7 (1 + g_A^2)}{96}\ ,
\eea
and the ones for the finite renormalization read
\bea
\bar{\beta}_m&=&-\frac{8g_A^2}{3}\IB\ ,\nonumber\\
\bar{\beta}_{\hat{c}_1}&=&\frac{g_A^2}{36}(1+\frac{\IB}{m^2})+\frac{g_A^2}{3m}(8\hat{c}_1+3c_7)(2m^2-3\IB)\ ,\nonumber\\
\bar{\beta}_{{c}_7}&=&\frac{5g_A^2}{12}(1+\frac{\IB}{m^2})-\frac{2g_A^2}{3}c_7m+c_7g_A^2\frac{\IB}{m}\ .
\eea

\end{document}